# Jump balls, rating falls, and elite status: A sensitivity analysis of three quarterback rating statistics

Laura A. Albert

University, of Wisconsin-Madison

Madison, Wisconsin, USA, 53706

Email: laura@engr.wisc.edu

John N. Angelis

Elizabethtown College

Elizabethtown, Pennsylvania, USA 17022

Email: angelisj@etown.edu

**Abstract**

Quarterback performance can be difficult to rank, and much effort has been spent in creating new rating systems. However, the input statistics for such ratings are subject to randomness and factors outside the quarterback's control. To investigate this variance, we perform a sensitivity analysis of three quarterback rating statistics: the Traditional 1971 rating by Smith, the Burke, and the Wages of Wins ratings. The comparisons are made at the team level for the 32 NFL teams from 2002-2015, thus giving each case an even 16 games. We compute quarterback ratings for each offense with 1-5 additional touchdowns, 1-5 fewer interceptions, 1-5 additional sacks, and a 1-5% increase in the passing completion rate. Our sensitivity analysis provides insight into whether an elite passing team could seem mediocre or vice versa based on random outcomes. The results indicate that the Traditional rating is the most sensitive statistic with respect to touchdowns, interceptions, and completions, whereas the Burke rating is most sensitive to sacks. The analysis suggests that team passing offense rankings are highly sensitive to aspects of football that are out of the quarterback's hands (e.g., deflected passes that lead to interceptions). Thus, on the margins, we show arguments about whether a specific quarterback has entered the elite or remains mediocre are irrelevant.

# 1 Introduction and Literature Review

A reoccurring sports challenge is how to compare players at a given position using one ranking statistic. Baseball sabermetricians have created many alternatives to the simple batting average statistic. Basketball has seen the rise of statistics such as Player Efficiency Rating (PER) rather than points (Hollinger, 2012). Quarterback (QB) is perhaps the most hotly debated position in sports, as fans and sports show hosts continually debate whether a given quarterback is elite or better than his rivals. Statistics such as team wins, touchdowns, passing yards, and interceptions have been used to further the debate, as have methods such as linear programming models (Erickson and Callum 2004) , tiered logistic regressions (White and Berry 2002), and ordinary least squares regressions (Stern 1998). The traditional QB Rating was devised by Smith in 1971. Since then, the Berri QB Score, also known as Wages of Wins rating (Berri et al. 2007), and the Burke method (Winston 2009), attempted to improve upon Smith's work.

For the most part, research into QB ratings has focused on how to build a better rating. Schatz (2005) called for more data to be collected, such as separating yards per completion into its catch and run-after-catch components, while Joyner (2005) plotted QB performance by zones (short, medium and deep) and tracked receivers and defenders as well. Since 2005, data specificity and depth has improved, including All-22 camera angles that allow for better player tracking. One would expect better systems to emerge for QB evaluation. Play-by-play QB ratings such as ESPN's Total Quarterback Rating system (Oliver 2011) have recently been created. Other ratings that take an in-depth look at the quarterback's performance include Cold Hard Football Facts (2016) and Football Outsiders (2016). In addition, some have returned to the Traditional QB Rating and proposed modest adjustments (von Dohlen 2011). The correlation between QB performance and winning shows the need for further study.

However, in this rush to build a better mousetrap, researchers may have overlooked the extreme sensitivity of rating systems to nearly random results. Consider, for example, the work of Berri and Burke (2012) in tracking year-to-year quarterback performance. Unlike basketball and baseball, where 40-60% of a player's current performance can be determined by his performance last season, only 15% of a quarterback's current rating can be determined by his last season's rating. In Berri and Burke's study of quarterbacks from 1998-2010, only 31.1% of completion percentage, 10.1% of touchdowns per attempt, and 0.6% of interceptions per attempt in a current season could be explained by the quarterback's previous season.

One hypothesis that Berri and Burke use to explain these results is that it is difficult to isolate the quarterback's performance from that of coaches, teammates, and defenses faced. For example, Alamar and Weinstein-Gould (2008), in studying offensive linemen, find that the loss of one offensive lineman cost his team 3% of their completed passes. The optimal ratio of passing to running plays often does not match the actual choices of playcallers, for reasons such as risk-aversion and under-appreciation of game-theoretic elements (Alamar 2006, Jordan et. al. 2009, McGough et. al. 2010, and Rockerbie 2009). Yet, we point out that free agent movement occurs less often in the NFL than in basketball and baseball, and that NFL coaches are on average tenured longer than baseball and basketball coaches (Dodd, 2012). And while football teams do not play a complete round-robin schedule, baseball and basketball teams play an unbalanced round-robin schedule.

However, an underappreciated issue is that non-quarterback elements of the passing game are predictable. The success of NFL wide receivers and tight ends (Mulholland and Jensen, 2014) correlates well with college statistics and combine variables. For example, they found that good NFL receivers either caught a lot of touchdowns in college or, to compensate, had great final years in college before going pro. Even with the evolution of tight ends from mainly blockers to more frequent pass catchers, Mulholland and Jensen found a surprising amount of consistency between the two eras.

However, when Wolfson et al. (2011) attempted a similar study on quarterbacks, they found that "College and combine statistics have little value." The problem was not that NFL teams sub-optimally collect qualitative and quantitative data. No, they concluded the problem was "random variability in future performance due to factors which are unlikely to be observable." We thus focus on that random variability in our research. Further support for our approach is found in Berri and Burke's second hypothesis for issues in year-to-year QB ratings: "Interceptions clearly have a big impact on outcomes. But a quarterback's interceptions are not predictable." We also note that Barnwell (2012) investigated Michael Vick's last 13 starts and found that more than 40% of Vick's interceptions were due to tipped balls (either by defenders at the line of scrimmage or his own receivers). These tipped balls represent jump balls of sorts that could easily be a completion, interception, or touchdown, for reasons that seem to be more based on randomness than skill. Given this randomness, and that interceptions serve as an input statistic in all quarterback ratings, it is worthwhile to investigate how sensitive QB ratings are to slight changes in a quarterback's interception total, as well as touchdowns, sacks, and completion percentages. Our work investigates this for the Smith, Burke, and Wages of Wins rating systems.

In addition, we compute to what extent these changes in QB ratings affect overall rankings of quarterbacks. Stimel (2009) analyzed NFL QB Rating variables and found that comparisons between eras may be inappropriate due to changes in input statistics. We limit ourselves to a decade's worth of data in order to avoid this. In addition, he goes on to show that input statistics may be causally related (e.g., completion percentage and interception percentage). Because we are doing sensitivity analysis, and because Stimel's method (graph theory) of determining causation does not give precise factors of causation, we do not make any causation inferences. Finally, Stimel hypothesizes that due to the improvement in quarterback performance over the years, the difference between above average and average quarterbacks has diminished. An advantage of our sensitivity analysis is that we can precisely measure how that difference changes due to fluctuation in input statistics.

We compute quarterback ratings for each offense in the base case and with 1-5 additional touchdowns, 1-5 fewer interceptions, 1-5 additional sacks, and a 1-5% increase in the passing completion rate (about 5-25 more completions per season). Note that these fluctuations are reasonable: 1 game out of 16 can account for five additional touchdowns, interceptions, or sacks, and 25 additional completions is approximately 1.5 more completions per game. We then compute the number of rank changes (e.g., rank change = 1 when a quarterback rises or falls by one position in the rankings) that occur based on the scenario. Even if quarterback performance were inherently consistent from year to year, outside changes create such fluctuations, such as the 3% change in completion rate reported by Alamar and Weinstein-Gould due to the loss of one talented lineman.

Our sensitivity analysis provides insight into whether an elite passing team could seem mediocre (or vice versa) based on random outcomes. Our comparisons are made at the team level (rather than at the player level) for the 32 NFL teams for two reasons. First, a team's passing offense is a good proxy for individual passer performance, since a single quarterback makes nearly every pass attempt for many NFL teams. Second, the QB ratings have different scales, and therefore, performing a sensitivity analysis on each rating does not necessarily shed light on the extent to which a quarterback (or passing offense in this case) changes its rank based on an extra touchdown pass, for example.

The results indicate that the Traditional rating is the most sensitive statistic with respect to touchdowns, interceptions, and completions, whereas the Burke rating is most sensitive to sacks. All ratings are sensitive to completions. On average, one additional touchdown results in 17.9 total rank changes in the Traditional rating per year, 2.7 in the Burke rating, and 3.9 in the Wages of Wins rating (across 32 teams). In other words, a QB has a greater than 50% chance to move up one spot in the rankings by tossing one more TD for the season. As context, note that in our data, the top 10 ranked QB's in TD's throw 1.5-3 TD's per game. The rank changes increase to 60.2, 13.1, and 21.3 in the Traditional, Burke, and Wages of Wins ratings, respectively, when there are five additional touchdowns. The analysis suggests that team passing offense rankings are highly sensitive to aspects of football that are out of the quarterback's hands (e.g., deflected passes that lead to interceptions. For example, five additional interceptions results in an average of 108.2 (3 spots for average team), 97.1, and 82.5 rank changes in the Traditional, Burke, and Wages of Wins ratings. Moreover, a 3% increase in the completion rate (just one more catch per game) results in an average of 56.4, 50.4, and 55.6 rank changes in the Traditional, Burke, and Wages of Wins ratings.

## 2 Methodology

Our methodology uses passing offense data from the 32 NFL teams from 2002-2015. The data used include the number of passing attempts (ATT), completions (COMP), passing yards (YDS), interceptions (INT), touchdown passes (TD), sacks (SK), sack yards lost (SKYD).

- One additional touchdown pass is evaluated as one extra completed pass and attempt (that is a touchdown) with the number of yards equal to the team's average yards per completion.
- One fewer interception is evaluated as one fewer attempted pass.
- One fewer sack is evaluated as one fewer offensive play. It does not affect the number of completed or attempted passes.
- A 1% increase in completed passes transforms existing incomplete passes into complete passes based on a team's number of attempts over the season. The additional number of yards associated with each completed pass is equal to the team's average number of yards per completion.

The three quarterback ratings are summarized in Winston (2009) and are as follows:

1. Traditional Rating: (100 / 6) x [ 5 ((COMP/ATT)-0.3) + 20 (TD/ATT) + (2.375 – 25 (INT / ATT)) + 0.25 ((YDS / ATT) – 3))]

2. Burke Rating:  1.543 (YDS – SKYD) / (ATT – SK) + 50.0957 (INT / ATT)

3. Wages of Wins Rating:  YDS – 3(ATT+SK) – 30(INT)

We examine the impact of 1-5 additional touchdowns (TD+1, TD+3, TD+5), 1-5 fewer interceptions (INT-1, INT-3, INT-5), 1-5 additional sacks (SK+1, SK+3, SK+5), and a 1-5% increase in the passing completion rate (Comp+1%, Comp+3%, Comp+5%) on QB rating rank changes.  To do so, we changed one team's rating at a time and computed the number of rank changes. Then, we summed the total number of rank changes across the 32 teams.  The distribution of rank changes was approximately normally distributed. Therefore, we performed Hypothesis tests using the Student T-distribution assuming a pooled variance to identify statistically significant differences in the sum of rank changes between the ratings across the 32 teams.

## 3  Results

Table 1 reports the average and standard deviation of the number of rank change per year (across 32 teams) according to the sensitivity analyses. The Burke rating, for example, results in 97.1 rank changes when considering five fewer interceptions, which indicates that a team can move up or down in the ranks by an average of three positions. The average number of rank changes per team is illustrated in Figure 1a. All three QB ratings are quite sensitive to interceptions and completions and somewhat less sensitive to touchdowns and sacks. The exception is the Traditional QB rating, which is highly sensitive to touchdown passes and completely insensitive to sacks.

Table 1: The average (standard deviation) number of rank changes per year across 32 teams.

| Sensitivity | Traditional | | Burke | | Wages of Wins | |
|---|---|---|---|---|---|---|
| TD+1 | 17.9 | ( 6.6 ) | 2.7 | ( 1.7 ) | 3.9 | ( 1.7 ) |
| TD+3 | 48.1 | ( 14.1 ) | 7.8 | ( 3.3 ) | 12.3 | ( 4.0 ) |
| TD+5 | 60.2 | ( 16.8 ) | 13.1 | ( 3.8 ) | 21.3 | ( 6.1 ) |
| INT-1 | 24.6 | ( 7.1 ) | 17.0 | ( 4.6 ) | 15.4 | ( 4.7 ) |
| INT-3 | 65.7 | ( 12.9 ) | 64.2 | ( 32.7 ) | 51.1 | ( 9.6 ) |
| INT-5 | 108.2 | ( 20.4 ) | 97.1 | ( 24.2 ) | 82.5 | ( 14.4 ) |
| SK+1 | 0.0 | ( 0.0 ) | 5.6 | ( 2.6 ) | 1.5 | ( 1.1 ) |
| SK+3 | 0.0 | ( 0.0 ) | 15.6 | ( 5.3 ) | 4.3 | ( 1.8 ) |
| SK+5 | 0.0 | ( 0.0 ) | 27.8 | ( 5.5 ) | 7.2 | ( 3.0 ) |
| Comp+1% | 21.7 | ( 6.9 ) | 15.7 | ( 4.8 ) | 18.4 | ( 5.7 ) |
| Comp+3% | 56.4 | ( 11.0 ) | 50.4 | ( 11.8 ) | 55.6 | ( 11.9 ) |
| Comp+5% | 91.7 | ( 15.6 ) | 84.8 | ( 16.7 ) | 90.9 | ( 17.9 ) |

Table 2 and Figure 1b show the maximum rank change per team according to different sensitivity analyses. The largest number of rank changes tends to occur in the INT-5 category: 12 across all 32 teams and 6 across the top 8 teams. We note that in actuality, of course, survivor biases occur, and a team whose interception rate soared would bench the offending quarterback or run the ball more. The top 8 teams represent the "elite" quarterbacks at the high end of the pass rating distribution, since most discussions about elite quarterbacks range from the top 8 to the top 3 players. The maximum rank changes are somewhat smaller across the top 8 teams (as opposed to across all 32 teams). However, these results suggest that a quarterback's elite status could be lost based on dropped passes (leading to fewer touchdowns and completions) and tipped balls (leading to interceptions).

Table 2: The maximum rank change per team according to different sensitivity analyses.

|  | Maximum rank change across 32 teams | | | Maximum rank change across top 8 teams | | |
|---|---|---|---|---|---|---|
| Sensitivity | Traditional | Burke | Wages of Wins | Traditional | Burke | Wages of Wins |
| TD+1 | 4 | 2 | 2 | 2 | 1 | 1 |
| TD+3 | 8 | 3 | 3 | 4 | 2 | 1 |
| TD+5 | 8 | 3 | 4 | 4 | 2 | 2 |
| INT-1 | 5 | 3 | 4 | 3 | 2 | 1 |
| INT-3 | 8 | 6 | 7 | 4 | 5 | 4 |
| INT-5 | 12 | 8 | 9 | 4 | 6 | 5 |
| SK+1 | 0 | 3 | 1 | 0 | 2 | 1 |
| SK+3 | 0 | 3 | 2 | 0 | 2 | 1 |
| SK+5 | 0 | 4 | 2 | 0 | 3 | 1 |
| Comp+1% | 4 | 3 | 4 | 3 | 2 | 2 |
| Comp+3% | 8 | 6 | 7 | 4 | 5 | 4 |
| Comp+5% | 10 | 7 | 10 | 4 | 6 | 5 |

The differences in the number of rank changes between QB ratings (as shown in Figure 1 and Tables 1 and 2) are often significant. Table 3 reports statistically significant rank changes at the 0.05 level using one-sided hypothesis tests between two population means. This indicates that the Traditional QB rating is more sensitive to touchdowns, interceptions, and completions, whereas the Burke rating is more sensitive to sacks. The Traditional QB rating is the least sensitive with respect to sacks, a difference that is statistically significant. However, this lack of sensitivity is a weakness of the Traditional QB rating, since it does not account for potential passing plays that were lost.

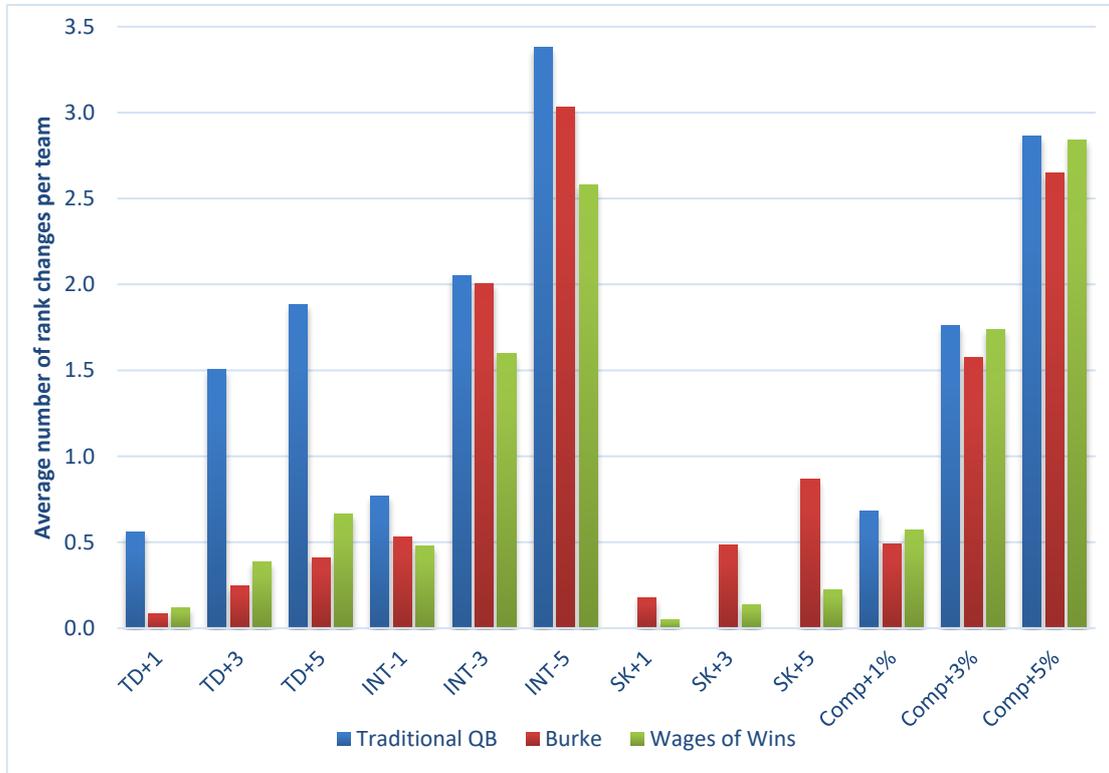

(a) average number of rank changes per team

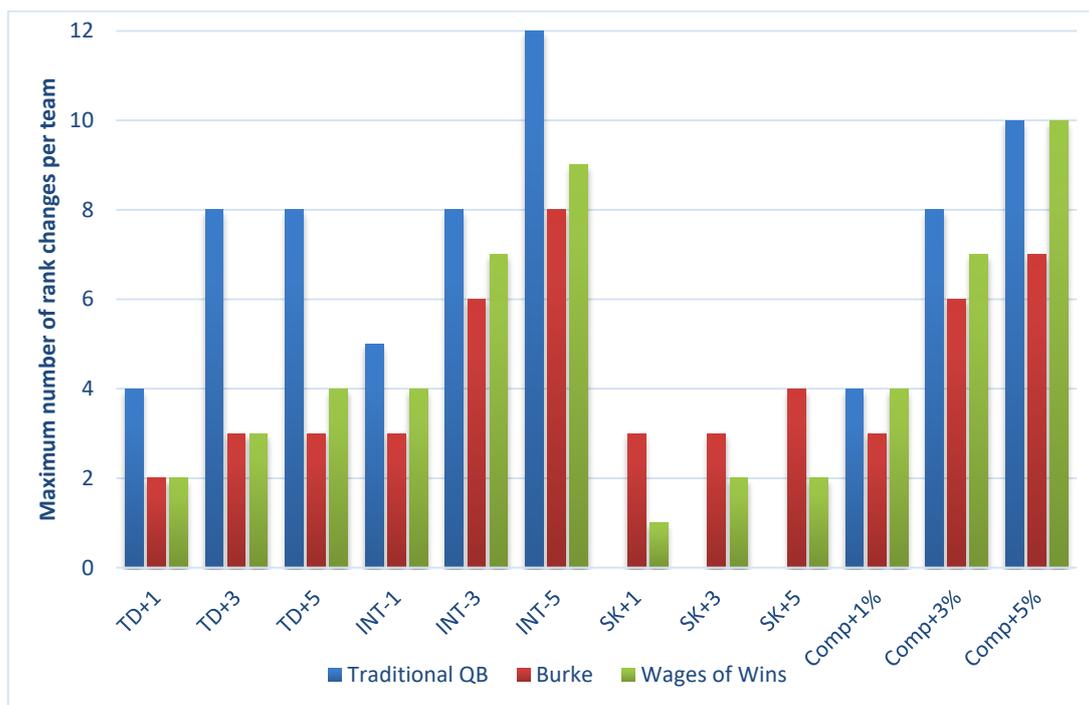

(b) maximum number of rank changes per team

Figure 1: The average and maximum number of rank changes per team according to different sensitivity analyses.

It is difficult to identify an "optimal" range for a QB rating's sensitivity. However, we can shed light on appropriate relationships across the different types of sensitivity analyses performed. Five sacks (6.75 yards lost per sack in 2011) is approximately equivalent to having lost five pass attempts (6.5 yards per attempt in 2011) but not as costly as losing five completions (11.5 yards/completion in 2011). A 1% increase in completions is approximately 5 passes. Therefore, a quarterback rating ideally should be approximately sensitive to 5 additional touchdown passes as it is to a 1% increase in completions (although we note that a completed touchdown pass yields 6-8 points as well as the additional yards), and 5 additional touchdown passes should be less sensitive to 5 additional sacks. We find that the Wages of Wins rating and the Burke rating to a lesser extent are consistent across touchdowns, completions, and sacks. The Traditional QB rating appears to be too sensitive with respect to touchdown passes and not sensitive enough with respect to sacks.

## 4 Case Studies and Discussion

Ultimately, different QB ratings are used to determine which quarterbacks are "elite," which in turn may influence offensive play calls, Pro Bowl selections, and quarterback salaries. We expand on the analysis of Table 2 on elite teams. To further explore the role of sensitivity and elite status, we perform a case study on how the ratings of the top 9 passing teams in the 2009 season change due to ±3-interceptions. Those teams had a mean of 12.1 interceptions and standard deviation of 4.1. Table 4 summarizes the results, where asterisks (*) indicate a non-elite ranking (7 or lower). We are using 6 because that seems safely in the range of QB's who are universally judged elite (i.e., top 3 QB's in both leagues are selected for Pro Bowl). Note that overall, positions in the top 9 change less than Figure 1a would predict. In particular, the top 4 teams are not greatly affected by fluctuations around their interception rate. The biggest top 4 team jump from decreased interceptions is #4 Dallas for Wages of Wins, moving a mere 2 places, and the biggest fall from increased interceptions is New Orleans from #2 to #5.

However, teams ranked 5 – 9 can move in and out of the elite category based on ±3 interceptions. For example, Indianapolis soars from 9$^{th}$ to 5$^{th}$ in the traditional ratings if 3 fewer interceptions are thrown, and in fact all teams ranked 6$^{th}$ or less via the Traditional QB rating can improve to 5th if 3 fewer interceptions are thrown. Significant but smaller changes are observed for the other two rating systems. This places our work in proper context. The sensitivity of QB Rating systems does not turn a Tom Brady into a mediocre quarterback, but it does make it difficult to accurately judge the ability of all but the best quarterbacks.

Table 3: Quarterback ratings with statistically significant rank changes (at the 0.05 level) using one-sided hypothesis tests between two population means.

| Sensitivity | Traditional vs. Burke | Traditional vs. Wages of Wins | Burke vs. Wages of Wins |
|---|---|---|---|
| TD+1 | Traditional QB | Traditional QB | - |
| TD+3 | Traditional QB | Traditional QB | Wages of Wins |
| TD+5 | Traditional QB | Traditional QB | Wages of Wins |
| INT-1 | Traditional QB | Traditional QB | - |
| INT-3 | Traditional QB | Traditional QB | - |
| INT-5 | Traditional QB | Traditional QB | - |
| SK+1 | Burke | Wages of Wins | Burke |
| SK+3 | Burke | Wages of Wins | Burke |
| SK+5 | Burke | Wages of Wins | Burke |
| Comp+1% | - | - | - |
| Comp+3% | - | - | - |
| Comp+5% | - | - | - |

We also examine two specific quarterbacks in that 2009 database. Table 5 reports Tony Romo's 2009 rating and his rank change with three additional interceptions. In 2009, the Dallas Cowboys were ranked 5$^{th}$, 3$^{rd}$, and 4$^{th}$ under the Traditional, Burke, and Wages of Wins rankings, respectively, and Romo played all 16 games. This season qualifies him as an elite quarterback. However, if he had thrown three extra interceptions, his rankings would fall to 9$^{th}$, 5$^{th}$, and 6th, respectively. This suggests that factors outside of a quarterback's control could lead to a quarterback losing his "elite" status and perhaps missing out on the Pro Bowl (the top 3 or so quarterbacks in each league).

Finally, we investigate the impact of ±3-interceptions on the Indianapolis Colts during the years 2005-2010, in which their quarterback, Peyton Manning, was judged to be elite by most observers. He threw an average of 13 interceptions over those years (high of 17, low of 9). The results are shown in Table 6, where asterisks (*) indicate a non-elite ranking (7 or lower) as before. His 2009 year would not be elite (i.e., top 6) by our standards, but otherwise his original rankings are all top 6 or better. If Manning threw 3 fewer interceptions each year, his elite status is never questioned. In 2005 and 2006, when he is ranked first, throwing 3 more interceptions only drops him to second, and that once. However, if he threw 3 more interceptions in 2008-2010, he is not ranked as elite, except for the Traditional rating in 2008 and Wages of Wins in 2010. We thus illustrate, using a top quarterback, that given the sensitivity of ranking systems, judging quarterbacks by a fixed number of elite slots is not recommended, even when the debate is limited to Pro Bowl level (i.e. top 6). There is too much potential variation due to randomness in the input statistics.

Table 4: Rankings for Top 9 Teams with ±3 Interceptions, 2009 (Asterisks Indicate a Ranking of 7 or Lower).

| Team, 2009 | Traditional | | | Burke | | | Wages of Wins | | |
|---|---|---|---|---|---|---|---|---|---|
| | INT-3 | **Base** | INT+3 | INT-3 | **Base** | INT+3 | INT-3 | **Base** | INT+3 |
| Minnesota | 1 | **1** | 2 | 4 | **6** | 7 | 4 | **6** | 8* |
| New Orleans | 1 | **2** | 2 | 2 | **2** | 2 | 2 | **2** | 5 |
| San Diego | 3 | **3** | 4 | 1 | **1** | 1 | 1 | **1** | 1 |
| Green Bay | 4 | **4** | 4 | 8* | **8*** | 8* | 5 | **7*** | 8* |
| Dallas | 5 | **5** | 9* | 3 | **3** | 5 | 2 | **4** | 6 |
| Pittsburgh | 5 | **6** | 9* | 9* | **10*** | 11* | 9* | **9*** | 10* |
| Houston | 5 | **7*** | 9* | 3 | **5** | 6 | 2 | **3** | 5 |
| New England | 5 | **8*** | 9* | 3 | **4** | 6 | 2 | **5** | 7* |
| Indianapolis | 5 | **9*** | 9* | 6 | **7*** | 7* | 6 | **8*** | 8* |

Table 5: Tony Romo's 2009 rank changes.

| Player | Traditional | | Burke | | Wages of Wins | |
|---|---|---|---|---|---|---|
| | Rating | Rank | Rating | Rank | Rating | Rank |
| Tony Romo (Base case) | 96.1 | 5 | 138.6 | 3 | 2265 | 4 |
| Tony Romo (INT+3) | 93.8 | 9 | 134.8 | 5 | 2175 | 6 |

Table 6: Peyton Manning's Rankings with ±3 Interceptions, 2005-2014 (Asterisks Indicate a Ranking of 7 or Lower)

| Year | Traditional | | | Burke | | | Wages of Wins | | |
|---|---|---|---|---|---|---|---|---|---|
| | INT-3 | **Base** | INT+3 | INT-3 | **Base** | INT+3 | INT-3 | **Base** | INT+3 |
| 2005 | 1 | **1** | 1 | 1 | **1** | 1 | 1 | **1** | 1 |
| 2006 | 1 | **1** | 1 | 1 | **1** | 1 | 1 | **1** | 2 |
| 2007 | 2 | **5** | 6 | 3 | **3** | 5 | 3 | **3** | 4 |
| 2008 | 3 | **5** | 5 | 5 | **6** | 8* | 4 | **5** | 7* |
| 2009 | 5 | **9*** | 9* | 6 | **7*** | 7* | 6 | **8*** | 8* |
| 2010 | 5 | **6** | 11* | 4 | **6** | 7* | 2 | **3** | 5 |
| 2012 | 1 | **2** | 2 | 1 | **1** | 1 | 1 | **2** | 3 |
| 2013 | 1 | **1** | 1 | 1 | **1** | 1 | 1 | **1** | 1 |
| 2014 | 3 | **4** | 4 | 3 | **3** | 3 | 2 | **4** | 4 |

## 5. Conclusion and Future Research

Our research sheds light on how random and unpredictable events outside of a quarterback's control can affect their ratings. Previous research has shown that a quarterback's touchdowns and interceptions show little correlation from year to year (Berri and Burke), and that completion percentage may be greatly affected by just one player's absence (Alamar and Weinstein-Gould), thus showing that many factors leading to touchdowns and interceptions are outside a quarterback's control.

The results indicate that all rating systems are quite sensitive to interceptions and completions and less sensitive to sacks, and that the Traditional rating is quite sensitive with respect to touchdowns. Thus, paradoxically, using QB rating statistics to separate quarterbacks with similar skill sets is inherently flawed, since slight fluctuations easily change QB rankings. The larger question, which remains open, is whether our results indicate that attempts to improve on the Traditional QB Rating system are more limited than previously thought. Even when data quality is fully improved (e.g., All-22 camera footage) and completions are better plotted and measured, that won't change the fact that touchdowns and interceptions are relatively rare events, and yet pivotal to most rating systems. There still is a need for a rating system that can separate luck from skill, and see past play call selection.

Another quarterback label worthy of further investigation, which we did not address, is the so-called "game manager" quarterback, who does not take many chances but is nonetheless successful. Current rating systems often struggle to identify the precise skill of such a quarterback, as the diminished number of both touchdowns and interceptions cancel each other out when compared to a less risk-averse quarterback. We hypothesize that if a rating system that is better attuned to sensitivity can be created, it would also rank quarterbacks better regardless of what role they are asked to play by their teams.